\newcommand{\ignore}[1]{}

\title{Daisen: A Framework for Visualizing Detailed GPU Execution}

\author[Y. Sun, Y. Zhang, A. Mosallaei, M. Shah, C. Dunne, \& D. Kaeli]
{\parbox{\textwidth}{\centering 
    Yifan Sun$^{1}$\orcid{0000-0003-3532-6521},
    Yixuan Zhang$^{2}$\orcid{0000-0002-7412-4669}, 
    Ali Mosallaei$^{3}$\orcid{0000-0001-9015-0018}, 
    Michael D. Shah$^{4}$\orcid{0000-0002-9093-133X},  
    Cody Dunne$^{4}$\orcid{0000-0002-1609-9776}, 
    and David Kaeli$^{4}$\orcid{0000-0002-5692-0151}
}\\
{\parbox{\textwidth}{\centering 
    $^1$ William \& Mary, 
    $^2$ Georgia Institute of Technology,
    $^3$ Lexington High School,
    $^4$ Northeastern University
}}}

\definecolor{colorAbstractTask}{HTML}{417505}

\DeclareRobustCommand{\colorRect}[1]{\tikz \draw [fill={#1},draw=none] (0,0) -- (0.4,0) -- (0.4,0.2) -- (0,0.2) -- cycle;}

\definecolor{colorEvent}{HTML}{5b85e2}
\definecolor{color_parent_task}{HTML}{A8A3E3}
\definecolor{color_current_task}{HTML}{A6DBB1}
\definecolor{color_annotation}{HTML}{D7191C}

\begin{document}

\teaser{%
    \centering%
    \includegraphics[width=\linewidth]{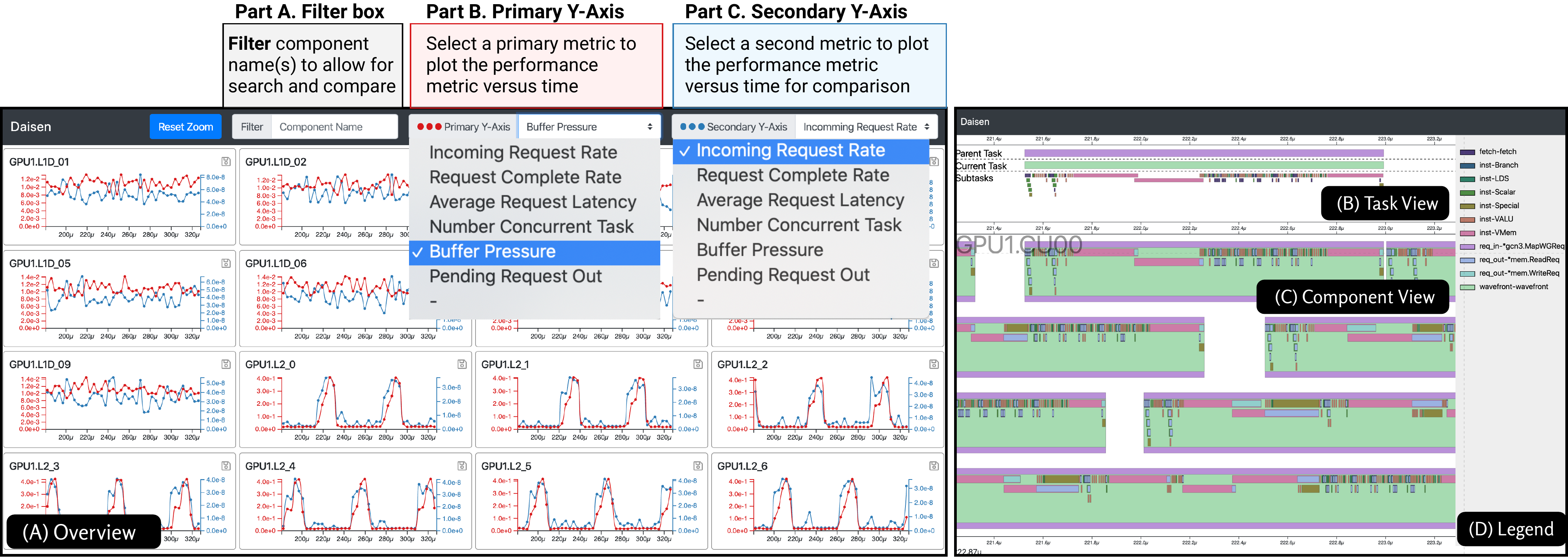}%
    \caption{Daisen provides a web-based interactive visualization tool that enables the examination of GPU execution traces. 
    (A) The Overview Panel shows key performance metrics against time of all hardware components using small multiples,  
    (B) the Task View demonstrates the hierarchical relationships between tasks, 
    (C) the Component View displays all the tasks executed in a hardware component (after selecting one component of interest from the Overview), and 
    (D) the legend bar shows all the tasks involved in the Component View and the Task View with color encoded.  
    Note that the labels ABCD do not appear on the visualization interface.}%
    \label{fig:teaser}%
}

\maketitle
\begin{abstract}
   Graphics Processing Units~(GPUs) have been widely used to accelerate artificial intelligence, physics simulation, medical imaging, and information visualization applications. To improve GPU performance, GPU hardware designers need to identify performance issues by inspecting a huge amount of simulator-generated traces. Visualizing the execution traces can reduce the cognitive burden of users and facilitate making sense of behaviors of GPU hardware components. In this paper, we first formalize the process of GPU performance analysis and characterize the design requirements of visualizing execution traces based on a survey study and interviews with GPU hardware designers. We contribute data and task abstraction for GPU performance analysis. Based on our task analysis, we propose Daisen, a framework that supports data collection from GPU simulators and provides visualization of the simulator-generated GPU execution traces. Daisen features a data abstraction and trace format that can record simulator-generated GPU execution traces. Daisen also includes a web-based visualization tool that helps GPU hardware designers examine GPU execution traces, identify performance bottlenecks, and verify performance improvement. Our qualitative evaluation with GPU hardware designers demonstrates that the design of Daisen reflects the typical workflow of GPU hardware designers. Using Daisen, participants were able to effectively identify potential performance bottlenecks and opportunities for performance improvement. The open-sourced implementation of Daisen can be found at \href{gitlab.com/akita/vis}{gitlab.com/akita/vis}. Supplemental materials including a demo video, survey questions, evaluation study guide, and post-study evaluation survey are available at \href{osf.io/j5ghq}{osf.io/j5ghq}.

\ccsdesc[500]{Computer systems organization~Single instruction, multiple data}
\ccsdesc[500]{Human-centered computing~Information visualization}

\printccsdesc   
\end{abstract}
\section{Introduction}
Graphics Processing Units~(GPUs) were originally designed to accelerate 3D graphics rendering. In the last decade, developers have found additional general-purpose use cases that take advantage of GPU's architectures. Developers have been using GPUs to accelerate a wide range of applications that require high performance, including large-scale physics simulation~\cite{navarro_hitschfeld-kahler_mateu_2014}, medical imaging~\cite{yan2008fast}, data visualization~\cite{McDonnelElmvvist}, artificial intelligence~\cite{oh2004gpu}, and blockchain hashing~\cite{morishima2018accelerating}. A GPU is capable of performing thousands of calculations in parallel. Empirical studies have demonstrated that GPU's massive parallel processing capabilities can speed up algorithms by 5-70X as compared to CPU implementations~\cite{AroraManish2012RtRo, sun2019summarizing}. 

GPU hardware designers aim to improve the design of a GPU to run applications faster. Prior to implementing a new design on real hardware and sending the design to fabrication, extensive simulations of the proposed design must be carried out.
Simulations allow hardware designers to compare the performance of the proposed designs with a baseline over a wide range of benchmarks (i.e., standard and representative applications)~\cite{gpgpusim, multi2sim, multi2sim_kepler, sun2019mgpusim, accelsim}. 
It is challenging to identify and make sense of performance bottlenecks encountered, since this may involve the designer manually inspecting gigabytes (GBs) of simulator-generated GPU-execution traces (i.e., highly detailed records of events that occur during GPU program execution)~\cite{M2S-Visual}. Given the complexity and the massive scale of the execution trace data, visualizations can help GPU hardware designers inspect execution states, identify potential performance issues, and improve the efficiency of the GPU hardware design process~\cite{aerialvision}. 

Existing performance analysis tools~\cite{gprof} typically assume the underlying execution data is organized using call graphs (a graph representation where the nodes are functions and the links are function calls). While call graphs are a reasonable representation for tracking software execution, they are not suitable for hardware execution. The problem is that a ``function'' is a software concept and is not used by hardware. Several visualization tools~\cite{isaacs2014combing, ross2019using, xie2018visual, fujiwara2017visual} have been developed to visualize parallel execution. However, these tools mainly focus on visualizing high-performance distributed computing and cannot be directly applied to hardware design.  
Visualization tools for examining GPU execution traces, such as AerialVision~\cite{aerialvision} and M2S-Visual~\cite{M2S-Visual}, only show specific metrics for specific features (e.g., instruction execution, cache access). Additionally, these tools lack a straightforward methodology to compare metrics when evaluating complex hardware tradeoffs.

We introduce Daisen, a framework that helps GPU hardware designers collect and visualize GPU execution traces. We contribute:

\begin{itemize}
    \item A task analysis and task abstraction of GPU performance analysis. To the best of our knowledge, there is no formal study that examines the design requirements of GPU execution trace visualization to facilitate GPU hardware design. To fill this gap, we conduct a survey and interviews with domain experts to characterize practices and workflows in GPU performance analysis.
    
    \item A data abstraction that is designed to capture simulator-generated traces of GPU executions.
    
    \item The design and implementation of Daisen, a web-based visualization tool that visualizes massively-parallel GPU execution traces.
    Our evaluation with domain experts demonstrates that Daisen can help GPU hardware designers identify performance bottlenecks, make informed and evidence-based design decisions, and efficiently debug performance issues.
\end{itemize}

\section{Background on How GPUs Work}

\begin{figure*}[t]
 \centering
 \includegraphics[width=\textwidth]{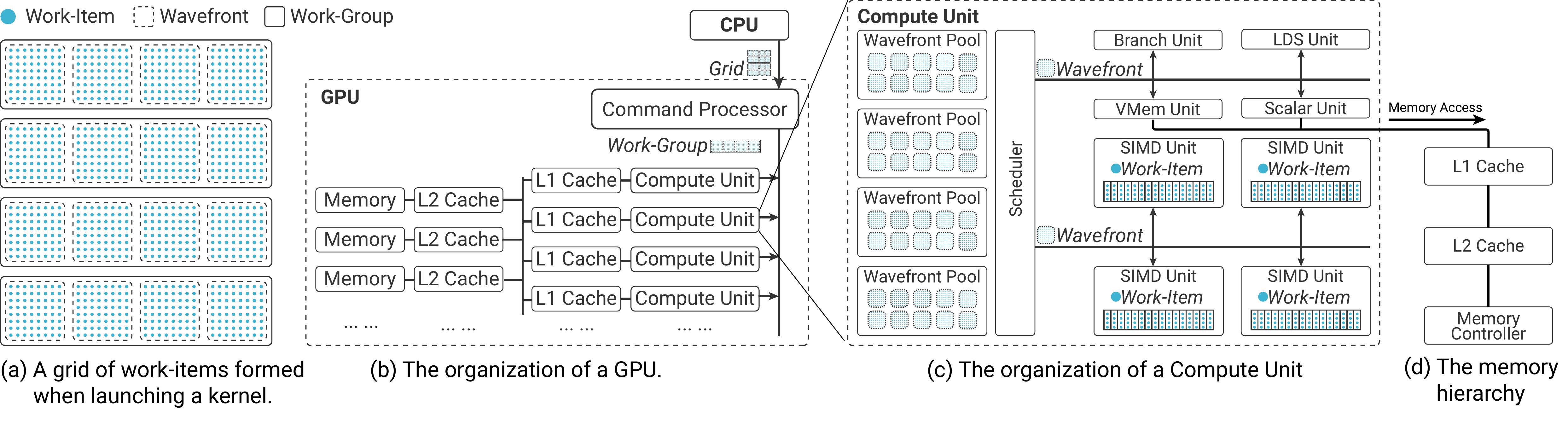} 
 \vspace{-1.0cm}
 \caption{An overview of the GPU execution model and hardware organization.}
 \label{fig:bg}
 \vspace{-0.6cm}
\end{figure*}

GPUs are computing devices that can perform a large number of data-processing operations concurrently. In this section, we briefly describe how GPUs and GPU simulators work, and summarize the challenges of visualizing highly parallel hardware architectures. 

\boldsec{GPU programming. }
GPUs work alongside CPUs in a coordinated effort to process data. A programmer writes a CPU program (i.e., host program) that prepares data and then copies the data over to the GPU to process.
After the required data is copied, the host program starts the GPU program (i.e., launching kernels).
GPU vendors usually provide programming APIs (e.g., CUDA, OpenCL) to perform memory copy and kernel launching tasks. 

A program that runs on a GPU is called a \textit{kernel}. As depicted in \autoref{fig:bg} (a), a kernel is composed of a large number of work-items that can be executed concurrently on a GPU. A \textit{work-item} is similar to a CPU thread that runs a series of instructions (e.g., add, multiply) to process data. Several work-items (typically 32 or 64, determined by the hardware) can be grouped into a \textit{wavefront}. Wavefronts (typically 1--8) are further grouped into a \textit{work-group}. 

Software elements (e.g., kernels, work-groups, wavefronts, work-items) need to be mapped to hardware resources for execution. We illustrate how the kernel is executed on a GPU with \autoref{fig:bg} (b) and (c). The CPU launches the whole kernel on a GPU for execution. On the GPU side, the \emph{Command Processor} receives the kernel launch request from the CPU and divides the kernel into work-groups. After this, the Command Processor dispatches the work-groups to the \emph{Compute Units}. Compute Units, similar to CPU cores, are responsible for calculating the program output. When work-groups are dispatched to a Compute Unit, the work-groups are broken down into wavefronts. Instructions that make up each wavefront wait to be scheduled by a central scheduler in each Compute Unit. The central scheduler dispatches instructions to the proper execution units (e.g., Branch Unit, SIMD unit), according to the type of each instruction. For a majority of the calculations, the instruction is sent to the \emph{SIMD units} (digital circuits that can process instructions from multiple work-items at the same time).

GPUs can process many work-items concurrently. E.g., the AMD R9 Nano GPU~\cite{r9nano} (default to MGPUSim and used in our case study and evaluation) has 64 Compute Units; each Compute Unit has 4 SIMD units, and each SIMD unit can calculate instructions from 16 work-items in one cycle. Thus, the whole GPU can calculate $64 \times 4 \times 16 = 4096$ instructions simultaneously. As the GPU runs at a 1 GHz frequency (1 nanosecond per cycle), it can execute 4 trillion instructions in 1 second with all the SIMD units. Given this volume of instructions that a GPU runs, it is essential to support visualizing this large dataset while still providing the capability of interactively viewing and zooming into the execution at an extremely fine granularity (e.g., sub-nanosecond level).

Moreover, Compute Units read data from memory (see \autoref{fig:bg} (d)). After data is copied from the CPU, it is stored in the GPU's main memory. Due to physics limitations, the main memory is too slow to feed the data required by the Compute Units. To increase throughput, GPU designers place faster, but smaller, storage close to Compute Units. Caches are organized in tiers, labeled as ``Level 1''~(L1) or ``Level 2''~(L2) caches, with the L1 caches being smallest, fastest, and placed closest to the Compute Unit. In many applications, the memory system implementation limits the performance of the whole GPU. Visually examining the memory system is crucial for understanding overall GPU performance.

\boldsec{GPU simulators. }
GPU simulators~\cite{gpgpusim, multi2sim, multi2sim_kepler, sun2019mgpusim, accelsim} recreate hardware behaviors with just a CPU. Hardware designers can run simulators and gather key performance metrics (e.g., execution time, number of instructions executed). A simulator can also dump details (i.e., traces) to a file to record all the events that occur during a simulation. Since the trace is highly detailed, simulating a millisecond GPU execution can generate several GBs of trace data.

\section{Related Work}

Performance visualization tools can help engineers to identify bottlenecks and make evidence-based performance-improvement decisions. Here we survey several such tools.

\boldsec{Call-graph visualizations. }
Many visualization tools for debugging performance issues in CPU programs rely on call graph data. These tools can be categorized into three groups: 
1. Icicle plots~\cite{kruskal1983icicle} or flame diagrams~\cite{gregg2016flame} are commonly used to show the hierarchical relationship between function calls and their execution time. Example tools include perf~\cite{PerfWiki77:online}, gperftools~\cite{gperftoo65:online}, callgrind~\cite{Valgrind35:online}, Chrome Developer Tools~\cite{chrome}, and Very Sleepy~\cite{VerySlee30:online}. 
2. Node-Link visualizations (e.g., pprof~\cite{pprofREA88:online}, kcachegrind~\cite{KCachegr71:online}) show function call relationships and use the size of nodes and links to visualize execution time. 
3. Pipeline-based visualization tools use multi-stage Gantt-chart~\cite{clark1922gantt} to show function executions (e.g., TraceVis~\cite{roberts2004tracevis}).
These tools were mainly developed for single-threaded CPU execution and cannot easily be used to visualize complex multi-threaded behavior.

Recently, new tools are developed to visualize multi-threaded executions. 
Isaacs et al.~\cite{isaacs2014combing, isaacs2015recovering} used logical times to align and compare the communication and dependency patterns between threads. Perfopticon~\cite{moritz2015perfopticon}, specifically designed for evaluating SQL query performance, could identify issues in query execution. Perfopticon used nested Gantt charts to represent the relationship between nested SQL queries.

Existing software-oriented performance visualization tools lack the proper abstractions needed to visualize GPU hardware execution. Hardware executions involve high degrees of parallelism, though without function calls, demanding new data abstractions and new visualization strategies. To address this research gap, our work aims to bridge event-based data representations (typically used to represent hardware execution) and call-graph-based data representations by building an abstraction of hardware behavior leveraging hierarchical tasks.

\boldsec{Visualizing parallel distributed computing. }
Most visualizations frameworks targeted parallel computing performance focus on large-scale distributed systems, and highlight the communication between computing nodes. Bhatele et al.~\cite{bhatele2012novel, isaacs2012exploring} developed Boxfish to visualize the multi-dimensional node organization on parallel systems, capturing the communication patterns between the nodes in a parallel distributed computing environment. Ross et al.~\cite{ross2019using} presented a tool that visualized not only the network traffic but also the behavior of the parallel simulation that models the communication. 
ChampVis~\cite{champvis} provided a solution to compare different hardware executions.
The capabilities of distributed-system network visualization have been extended recently with automatic route suggestions~\cite{fujiwara2017visual} and anomaly detection~\cite{xie2018visual, beschastnikh2020visualizing}.
However, hardware visualization poses different challenges (e.g., sub-nanosecond visualization, large number of component types), and hence, existing solutions cannot be directly used.


\boldsec{GPU performance visualization. }
GPU vendors have created profilers to enable developers to examine GPU performance (e.g., the Radeon GPU Profiler~\cite{radeon_gpu_profiler} and the NVIDIA Visual Profiler~\cite{profiler2014nvidia}).
Academia has also developed visualization tools to understand CPU-GPU heterogeneous execution~\cite{garcia2018visual}.
However, these tools mainly focus on high-level performance metrics in GPU software development and cannot show sub-nanosecond hardware execution details. Due to the lack of detailed GPU execution information, it is challenging for hardware designers to use industrial profilers to examine hardware behaviors.

There are a few tools developed specifically for visualizing GPU hardware execution.
Rosen~\cite{rosen2013visual} proposed a solution to visually inspect how GPUs access shared and global memory during CUDA kernel executions.
AerialVision~\cite{aerialvision} provides a time-lapse view for tracking the changes of several performance metrics. It presents a view of the statistics associated with individual lines of source code to help identify the root cause of bottlenecks. M2S-Visual~\cite{M2S-Visual} was customized to visualize execution traces generated by Multi2Sim~\cite{multi2sim}.
However, these existing tools lack the ability to easily compare metrics across hardware components. New visualization solutions are desired to simplify GPU performance analysis.

\section{Task Analysis and Abstraction}

\begin{figure*}[tb]
 \centering
 \includegraphics[width=\textwidth]{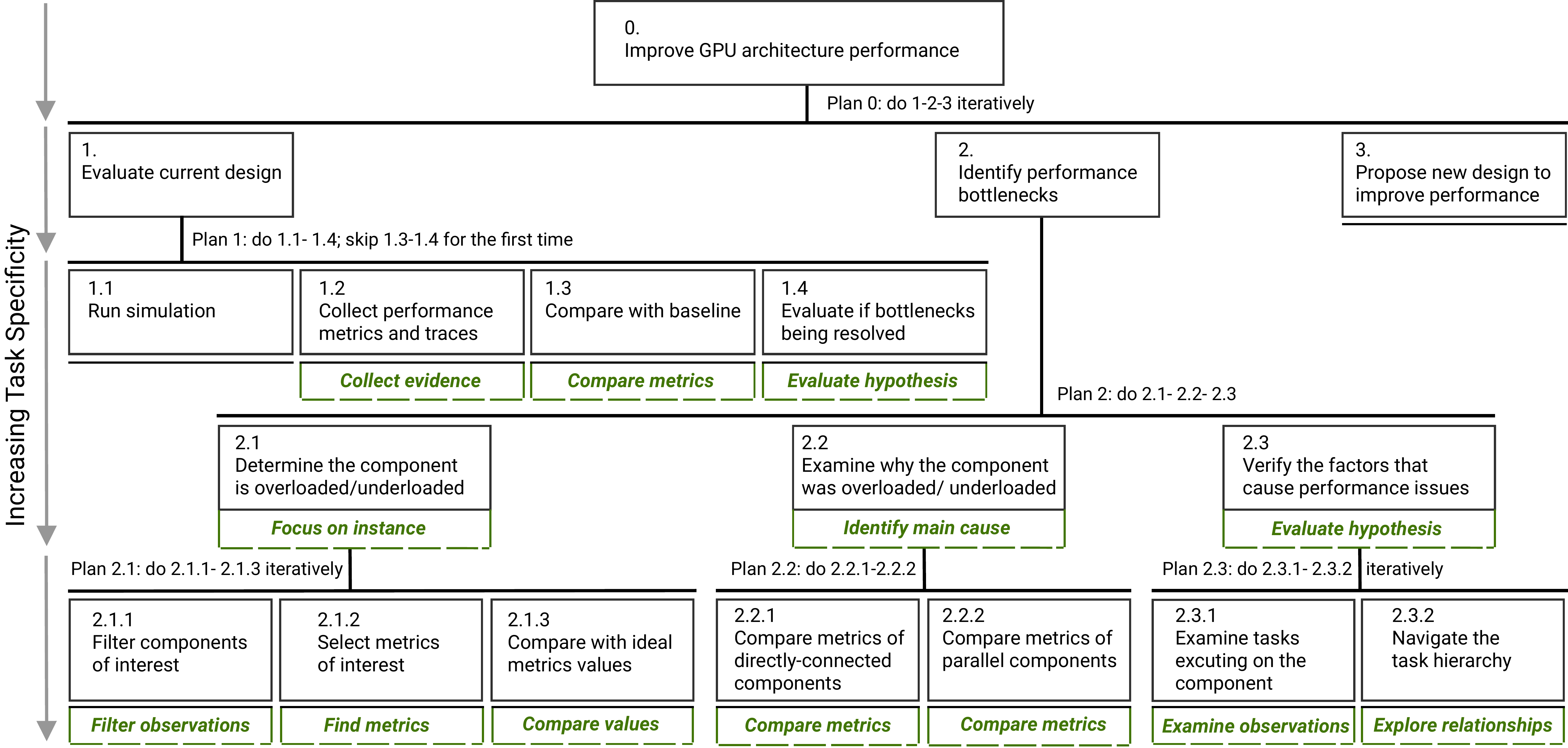} 
 \caption{Hierarchical Task Abstraction~\cite{zhang2019idmvis} of GPU hardware performance analysis: Each task is shown as a box with an ID that indicates the associated level of the hierarchy. The vertical lines connect the tasks and their subtasks. Labels on the vertical lines (i.e., plan) are used to identify the task sequences. Horizontal lines that sit beneath the corresponding boxes mean that there are no subsequent subtasks. Task abstractions derived from the tasks are highlighted in dashed green boxes \includegraphics[width=50pt]{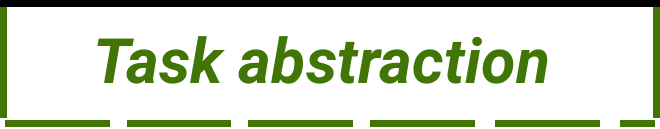}.
 }
 \label{fig:hta}
 \vspace{-0.6cm}
\end{figure*}

To characterize the workflow of GPU system performance analysis, we conducted a survey study and interviewed 5 GPU hardware designers (none of them are authors of this paper, upon approval from our Institutional Review Board). Below, we describe the processes of the survey study, interview sessions, and the domain characterization including data, task analysis and abstraction.

\boldsec{Study design and participants. }
We conducted a survey study with 37 computer hardware designers (none of them are authors of this paper) with experience analyzing parallel computing system performance. We created the survey using Google Forms (see supplemental materials) which was distributed via social media (e.g., Twitter, Slack). The survey contained three sections. \textbf{The first section} aimed to understand users' prior experience with performance analysis, such as how often respondents analyze computing system performance and use simulators during their daily work, what level of difficulty they experience when identifying and making sense of performance issues, and how much they feel that they need tools to help them examine performance issues. This section contained seven items using a five-point Likert-style scale, with choices ranging from (1) Strongly Disagree to (5) Strongly Agree.
\textbf{The second section} aimed to help us characterize workflows and strategies. Four open-ended questions were asked to understand what simulators they have used and the tools that help them identify performance issues, and to examine their process for identifying performance issues. \textbf{The last section} included five questions regarding respondents' positions and affiliations, length of working on performance analysis, and whether or not they would be willing to be contacted for clarification questions and follow-up studies.
Participants had various working experience on performance analysis: <1 year (n=7), 1--2 years (n=6), 2--3 years (n=6), 3--5 years (n=9), 5--10 years (n=7), and >10 years (n=2).

We also conducted semi-structured interviews with 5 GPU hardware designers to understand their daily practices of performing GPU performance analysis. The goal was to further refine the design requirements and characterize practices and common workflows when using simulators. The first two authors open-coded~\cite{thomas2006general} the survey responses and interview transcripts, and identified themes iteratively through discussion and affinity diagramming that helped us organize the codes.  

The aim of the survey and interview was to examine the workflow of carrying out performance analysis and guide us to help characterize domain tasks and processes. However, we are aware that the survey results may have potential biases since we used convenience sampling for recruitment (e.g., under-representation of sub-populations in the sample). Therefore, we avoid making absolute claims when analyzing and reporting our survey results, necessitating care when interpreting the results.

\boldsec{Task analysis. }
Based on the survey results and interviews, we performed a task analysis to characterize domain problems followed by task abstraction that aims to recast user goals from domain-specific languages to a generalized terminology for better understanding and generalizability~\cite{munzner2014visualization} (see \autoref{fig:hta}). We chose a Hierarchical Task Abstraction~\cite{zhang2019idmvis} as performance analysis in hardware design is a dynamic process that typically involves multiple hierarchical abstractions. 

The overall goal of GPU hardware designers is to improve GPU performance by proposing new GPU hardware designs (Task 0). The first step is to evaluate the current design's performance (Task 1). In the first iteration, designers evaluate the performance of the baseline design by running a simulation (Task 1.1) and collecting performance metrics and GPU execution traces (Task 1.2). 

After collecting the execution traces, GPU hardware designers need to identify performance bottlenecks (Task 2), checking if the components are overloaded or underloaded (Task 2.1). 
To complete this, designers need to find the component (Task 2.1.1), select the metrics to examine (Task 2.1.2), and compare with anticipated values (Task 2.1.3). 
After identifying the load on a component, the designer needs to find the root cause for each underloaded or overloaded component (Task 2.2) by comparing directly connected components (Tasks 2.2.1) and parallel components (Task 2.2.2).

To explore a hypothetical performance bottleneck, the GPU hardware designer typically examines execution details during a short time period (usually under 0.1 microsecond, Task 2.3).
The designer will first check the tasks that are executing in a particular hardware component (Task 2.3.1).
For example, they check all the instructions executing in a Compute Unit to see which instructions are preventing the Compute Unit from making further progress.
They also check the task hierarchy (Task 2.3.2). For example, a designer may want to check which memory accesses are conflicting with a memory load instruction.

After identifying the bottleneck of the current platform, the GPU hardware designers need to propose new designs that can mitigate the performance bottlenecks (Task 3). This step mainly relies on problem-solving skills and prior experience, thus it is out of the scope of this work. After developing a new design, designers need to repeat Task 1. This time, however, they need to compare the result with the baseline design (Task 1.3) and examine whether the bottleneck is resolved as expected (Task 1.4). If the performance of the new design cannot reach the performance goal, designers will look for a new performance bottleneck in the new design (Task 2). Designers may need to repeat Tasks 1--3 many times until they find a solution that satisfies the design requirements. Due to the page limitation, more detailed task analysis with examples are relegated to supplemental materials. 

\textbf{Task abstraction.}
Based on the task analysis, we performed a task abstraction that aims to recast tasks from domain-specific languages to a generalized visualization terminology to achieve better understanding and readability~\cite{munzner2014visualization}. We generate task abstractions guided by Amar et al.'s~\cite{amar2005low} low-level components of visual analytics and Lam et al.'s framework~\cite{lam2018bridging} that focuses on connecting analysis goals and steps in design studies. The abstractions, labeled green in~\autoref{fig:hta}, are integrated with the Hierarchical Task Abstraction guided by Zhang et al.'s work~\cite{zhang2019idmvis}.    

\boldsec{Design requirements. }
We identified a set of design requirements~(DR) of GPU execution trace visualization based on the task analysis.  
\textbf{DR1}: \textit{Support visualizing a large number of concurrently-executing tasks.} Hardware elements typically achieve a high degree of parallelism. Thousands of components may execute different tasks at the same time. Each component may also execute a large number of tasks in parallel.
\textbf{DR2}: \textit{Create a mapping between software concepts and hardware concepts.} It is essential to examine how the software elements (e.g., memory access) are executed on hardware components (e.g., caches). If designers question why a software element is not executed early enough, they need to investigate why the hardware elements are busy. 
\textbf{DR3}: \textit{Allow for efficient browsing through the execution and zooming down to focus on a short duration gradually.} Simulation traces record executions that last from milliseconds to seconds. Meanwhile, researchers need to identify hardware behavior at a sub-nanosecond level. 


\section{Data Abstraction}

\begin{table}[tb]
  \caption{A summary of data abstraction}
  \label{tab:data_abstraction}  
  \scriptsize%
  \centering%
  \begin{tabular}{@{}llp{4.2cm}p{1.6cm}@{}}
  \toprule
  \thead{Field} & \thead{Format} & \thead{Description} & \thead{Example}\\
  \midrule
  ID & Text & The unique identifier of a task, which was randomly generated to guarantee uniqueness & \texttt{5C9dX8} \\
  Parent ID & Text & The ID of the parent task & \texttt{7F3sY2} \\
  Category & Text & The category of the task belongs to & \texttt{Instruction} \\
  Action & Text & The job of the task & \texttt{Read Memory} \\
  Location & Text & The hardware component that carries out the task & \texttt{CPU1.Core1} \\
  Start/ End &  Time & The time that a hardware component starts to process/ completes processing the task & 0.00014566s\\
  \bottomrule
  \vspace{-0.8cm}
  \end{tabular}%
\end{table}

We first map data into a unified format. The goal is to build a cohesive and comprehensive data collection framework to help users make sense of the volumes of data and identify performance issues. We elect not to use prior parallel computing log formats~\cite{knupfer2006introducing, knupfer2008vampir, chan2000scalable} because: (1) Existing formats trace events rather than tasks. Pairing start-task events with end-task events requires extra processing. (2) It is difficult to capture the hierarchical relationship of the tasks with exiting formats. (3) While the design of prior formats considered query performance, it is challenging to use when working with file-based traces. Using a database program can address this issue and provide low-latency query services for a wide range to query types.

We summarize all the trace entities as tasks. Each task contains multiple properties, including ID, parent ID, category, action, location, start/end time, and details (see~\autoref{tab:data_abstraction}). Each task is assigned a unique ID. A task is also assigned information describing which category it belongs to and the job of the task, captured in the \textit{``Category''} (e.g., ``Request Out'', ``Request In'', ``Instruction'') and \textit{``Action''} (e.g., ``Read Memory'', ``Execute ADD Instruction'') fields, respectively. The name of the component that carries out the task is recorded in the \emph{``Location''} field. 
Each task has a \emph{Start} time and an \emph{End} time. 
Tasks are organized hierarchically using a tree structure. Each task has a \emph{``Parent ID''} field that tracks a task's parent task, except for the root task representing the entire simulation.

We record component-to-component communication with two special task categories: (i) ``Request Out'', and (ii) ``Request In''. Suppose a Compute Unit wants to read data from an L1 cache, the Compute Unit would initiate a ``Request Out'' task when the memory read request is created. When the L1 cache receives the request, a ``Request In'' task is created, setting the ``Request Out'' task as the parent task. After the L1 cache reads the data, either from its local storage or from an L2 cache, the L1 cache sends the data back to the Compute Unit and ends the ``Request In'' task. Finally, when the Compute Unit receives the response, the Compute Unit ends the ``Request out'' tasks. By tracking the ``Request In'' and the ``Request Out'' tasks, we can fully capture the communication between the components.

\section{Visualization Tool}

Next, we introduce Daisen's visualization tool, an open-source, web-based, interactive visualization tool that aims to help GPU hardware designers evaluate the performance of GPU designs. We highlight its capability of fulfilling the requirements with three unique perspectives, including the Overview Panel, the Component View, and the Task View.

\begin{figure}[tb]
 \centering
 \includegraphics[width=\linewidth]{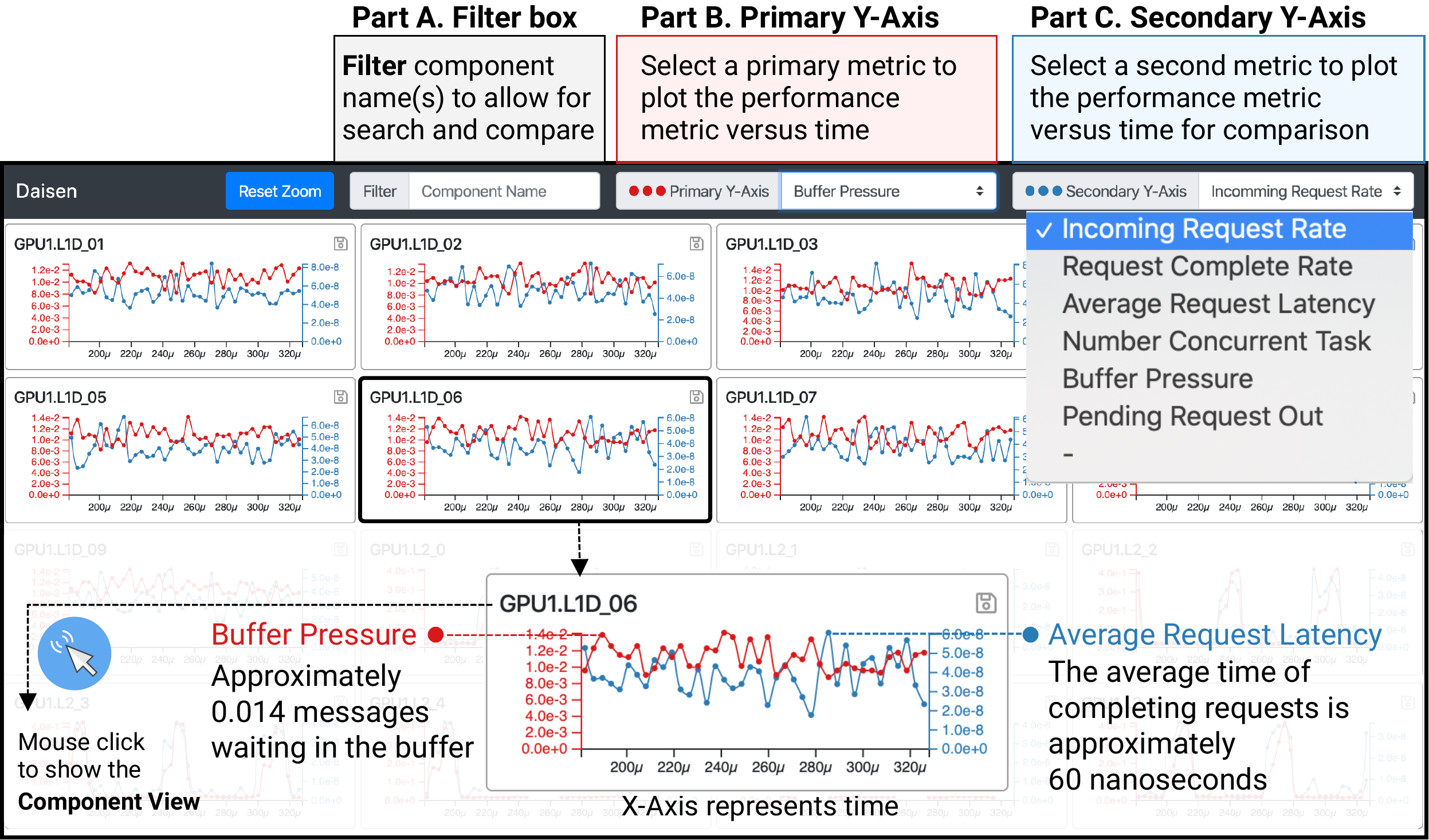}
 \caption{The Overview Panel shows multi-dimensional data. Users can filter components (Part A) using regular expressions and customize the display of dual y-axes (Parts B\&C).} 
 \label{fig:overview_view}
 \vspace{-0.4cm}
\end{figure}

\subsection{Overview Panel}
The Overview Panel (see \autoref{fig:overview_view}) enables users to see the big picture of the execution and locate interesting time intervals that require closer investigation --- partially fulfilling \textbf{DR-3}. 
The Overview Panel displays a series of small multiples, with one diagram per component. 
We separate the large number of GPU components into multiple pages.

We provide a \textit{filter box} (see~\autoref{fig:overview_view} Part A) to enable users to filter the components of interest, satisfying the requirement of targeted expert users. The filter box also supports task 2.2 (compare metrics of related components). Users can use regular expressions (e.g., \texttt{"(CU|L1|L2)"}) to create the combination of the components to compare.

We display all the component diagrams as time-series visualizations, with the x-axis indicating the simulation time. Users can select key metrics of interest to explore using a drop-down menu. They can also choose the metrics for both the primary y-axis (see~\autoref{fig:overview_view} Part B) and the secondary y-axis (see~\autoref{fig:overview_view} Part C).
We currently provide 6 metrics for the y-axes, selected according to the survey responses and feedback from domain experts.
Each data point represents the average value for the metric over a small time window. Although averages may hide outliers, they are effective for identifying general trends, which is in line with the goal of the Overview Panel.

When a user zooms or drags in a diagram, the time axes of all the diagrams will be updated to show the curves of the same time period. This creates an alignment effect and can help users compare the behavior of different components~\cite{zhang2019evaluating}.
Once the user arrives at an initial guess for the cause of the performance bottleneck, they may need more detailed information to verify the performance bottleneck, as in Task 2.3. The user can first zoom into a short time period in the Overview Panel and click the component name to see the tasks executing during that period in the Component View.

\begin{figure}[tb]
 \centering
 \includegraphics[width=\linewidth]{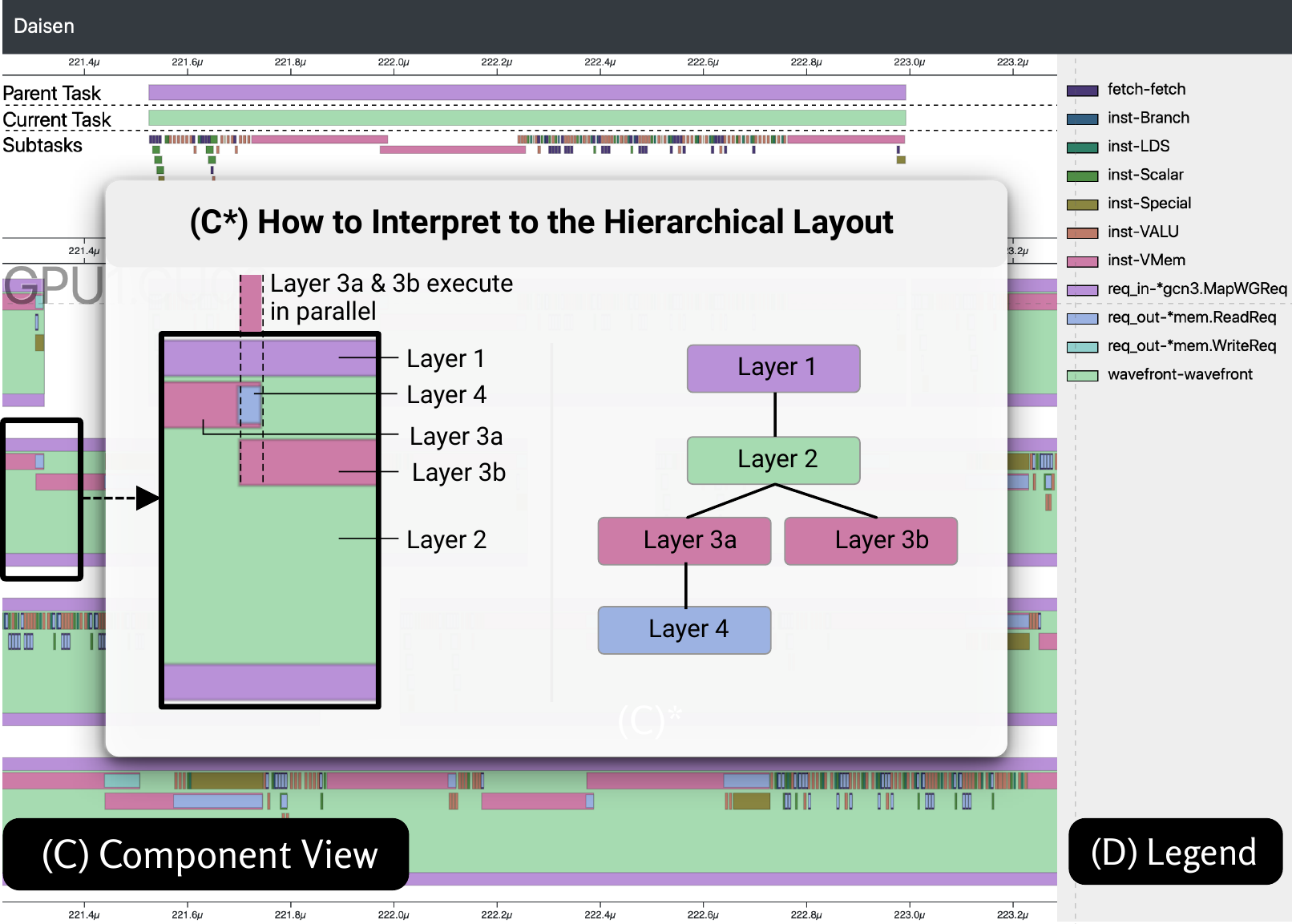}
 \caption{The Component View (C) hierarchically shows all the tasks executed in a given component. (C*) illustrates how the layout in the Component View maps to the tree structure. The legend (D) shows the color encoding for the tasks, 
 following the format of Category-Action. (Note: C* is not displayed in the interface.)}
 \label{fig:componenet_view}
 \vspace{-0.6cm}
\end{figure}

\subsection{Component View}

The Component View, as shown in \autoref{fig:teaser}(C), shows all the tasks executed at a certain component. The Component View aims to help users understand how hardware resources are utilized. The current component name (using transparent text) on the top-left corner of the Component View helps remind users which component they are currently viewing. 

\boldsec{Horizontal placement.} We use Gantt-like charts to represent tasks in the Component View, which is consistent with existing trace visualization tools~\cite{roberts2004tracevis, chrome}, flattening the learning curve and reducing mental burden~\cite{RogersYvonne2011IDbh}.

We choose to present the task visually in a hierarchical fashion to avoid clutter and allowing users to clearly understand the relationship between the tasks (see \autoref{fig:componenet_view}). In the Component View, root tasks (a task that has no parent task or its parent tasks have a different location) are directly displayed on the canvas. Each task uses its internal space to show its subtasks.

\boldsec{Vertical placement.} We develop an up-floating layout algorithm to determine where to place the task bars on the y-axis. We assign the row number for each task in the order of the task start-time. Each task is assigned with the smallest row number (top-most row), as long as it does not overlap with any task that is already assigned with that row number. In this way, we can use the minimum number of rows and show the maximum height possible for each task, while guaranteeing that tasks do not overlap with each other. 

We maximize the bar height to best utilize the vertical space. However, the bar heights are constrained by two conditions. First, no two bars should overlap to ensure readability. Second, we regulate the bar height so that same-level bars have the same height, to maintain logical consistency.

Users can rely on the number of rectangles stacked at the same x-position, indicating how many tasks are executing in parallel, on a specific component, at a particular time. This shows how busy the component is. The blank spaces on the y-axis are results of the bar height regularization, and should not be misinterpreted as the component being underutilized. We allow y-axis scrolling if there are too many tasks that are executing concurrently.

\boldsec{Color coding.} We use color-coding to differentiate tasks based on their ``Category'' and ``Action''. Since we cannot exhaustively list all the possible ``Category'' and ``Action'' values a priori,  we need to dynamically color the tasks. To solve this problem, we use the Cubehelix~\cite{cubehelix} coloring scheme, as  Cubehelix creates contrast in both the hue and the lightness, ensuring readability and accessibility for colorblind users. When there are too many ``Category''--``Action'' pairs, we fall back to code ``Categories'' to avoid using too many colors. 

One drawback of using a dynamic color-coding solution is that the colors can change as the user interacts with the tool.
E.g., when a user zooms out, more tasks will appear on the screen, using more ``Category''-``Action'' pairs. To differentiate the tasks, we need to assign more colors to the tasks.
The current recoloring solution changes the colors of tasks that are already on the screen and may add extra mental burdens for users.
A possible solution is to lock the color of the tasks already on the screen and generate new colors for newly appeared ``Categories'' and ``Actions''.
We will leave improving the dynamic color-coding solution as future work.


\boldsec{Interactive legend.} We reserve the right sidebar of the interface to provide information that may help users understand the diagrams. By default, color-coded information is displayed as a legend. The legend also allows users to find tasks of a particular type easily. When a user hovers the mouse pointer on a legend element, the tasks in the Component View, as well as the tasks in the Task View, will be highlighted with bolder outlines, while the rest is grayed out. The right sidebar can also provide additional information for each individual task. When the user hovers their mouse pointer on a particular task in the Component View, the detailed information (listed in \autoref{tab:data_abstraction}) will be shown above the legend.


\subsection{Task View}

The Component View demonstrates how the hardware is utilized. It aligns with a typical hardware performance analysis scenario. However, users also need to analyze the performance using a software performance analysis approach. They usually need to answer the question about why a particular task takes such a long time or which subtask takes the longest portion of the total time. Although the Component View hierarchically displays tasks, it cannot show subtasks that are not executed in the current task, and hence, does not allow the user to move up or down the entire task hierarchy, as required in Task 2.3.2. To solve these problems, we supplement the Component View with a Task View. A user can click on a task in the Component View to enable the Task View. The selected task becomes the ``current task''.

As shown in~\autoref{fig:teaser} (B), the Task View shows three groups of tasks. The second row of the Task View shows the ``current task''. Above the current task is its parent task. We show the parent task for two purposes. First, users can understand the relative timing relationship between the current task and the parent task, answering the question of whether the current task is a major step  required to complete the parent task. Second, users can click on the parent task to navigate towards the root of the task hierarchy, getting a larger picture of program execution and partially fulfilling \textbf{DR-3}.
All the subtasks of the current task are displayed below the current task. We reuse the ``up-floating'' algorithm to maintain consistency with the Component View and to save space. Showing the subtasks allows users to discover how the time is spent to complete the current task. Users can click on one of the subtasks to set that task as the ``current task'' or to drill down for a more detailed view. Combined with the parent task, the subtasks support navigation up and down through the entire task tree, fulfilling \textbf{DR-3}.

The Task View is stacked on top of the Component View. Their time axes are always aligned. Dragging or zooming one view always refreshes both views. Aligning the time axes helps the user to easily match the tasks in the Task View (software concept) and the Component View (hardware concept). When a user hovers over a task in either the Task View or the Component View, the task is highlighted on both sides to help users establish the connection between these two views. A typical use case is that a user may need to know why a task is not executed immediately after the previous task is completed. To answer this question, the user can check the task that occupies the component earlier. These two tasks are competing for hardware resources. As we align the Task View and the Component View, it should be straightforward to find the tasks that the component is executing at the time. The alignment of the Task View and the Component View can fulfill \textbf{DR-2}. 


\subsection{Implementation}
We provide a data collection library --- an extremely light-weight one with only 7 functions defined --- that can be invoked from the simulator to export trace data. 
Minimum changes are required to record the tasks. For example, we instrumented the MGPUSim's L1 cache model by adding 16 lines of code, as compared to the $\approx$3500 lines of code of the original L1 cache model. We have fully instrumented MGPUSim~\cite{sun2019mgpusim} and our instrumentation is a standard part of MGPUSim releases. We select to instrument MGPUSim because it is flexible and is widely used in state-of-the-art GPU architecture research projects~\cite{baruah2020valkyrie, mojumder2020mgpu, mojumder2020halcone, baruah2020griffin, li2019priority, tavana2019exploiting}.


The visualization tool is implemented with the client-server model. Users can start the backend server during the simulation execution or after the simulation completes. The server application reads data from the database, processes the data, and sends the data to the frontend upon request. The frontend was implemented using TypeScript and D3.js. This server-client model does not require the installation of any library or software except for a web browser on the client side. The server-client model also enables multiple users to check the results at the same time.

\section{Case Study}

\begin{figure*}[tb]
 \centering
 \includegraphics[width=\linewidth]{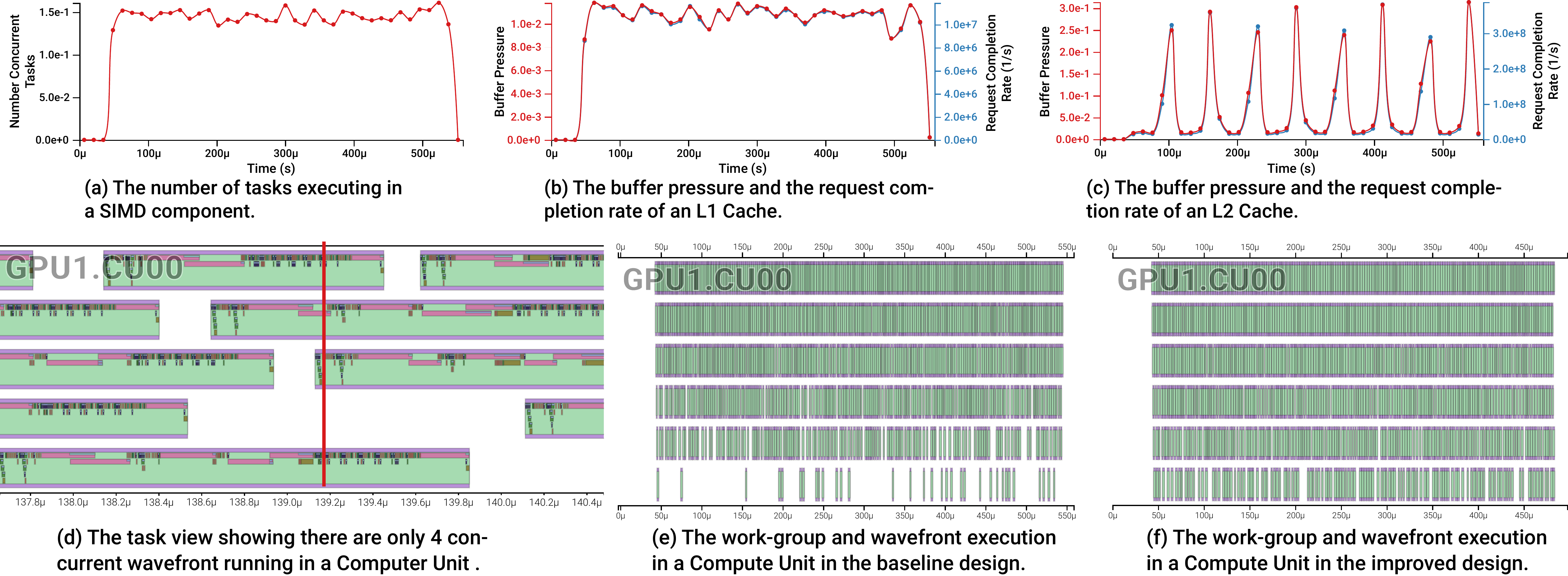}
 \caption{Example screenshots that help with workload characterization in the case study. }
 \label{fig:case_study}
 \vspace{-0.8cm}
\end{figure*}

We use a case study performed by a domain expert (an author of this paper)  to demonstrate how Daisen can help improve the performance of a GPU architecture design. We simulated the execution of the PageRank benchmark from the Hetero-Mark benchmark suite~\cite{sun2016hetero} running on an AMD R9 Nano GPU~\cite{r9nano}. 
We first ran the PageRank benchmark~\cite{pagerank} on an R9 Nano GPU with MGPUSim. MGPUSim reported the program could run in $\approx 51ms$. The simulation generated a trace with $\approx 32M$ tasks, consuming $\approx 6.1GB$ disk space.

The first task was to check which component was underloaded or overloaded to examine if the workload was compute-intensive or memory-intensive. To check compute intensity, we first checked the utilization of the SIMD unit (digital circuits that run the calculations) by filtering the SIMD unit by entering ``SIMD'' in the filter box and choosing ``Number of Tasks'' as the primary y-axis in the Overview Panel. According to \autoref{fig:case_study}(a), the SIMD utilization was low, suggesting it was not a compute-intensive workload. We then checked the ``Buffer Pressure'' and the ``Request Completion Rate'' for the L1 and L2 caches. From \autoref{fig:case_study}(b) and (c), we were able to conclude that both the buffers of the L1 and L2 caches were empty for most of the time and the Buffer Pressure value was constantly smaller than 1. We can conclude that this work was not memory intensive either.
Since the L1 cache was not given enough tasks (as suggested by the low buffer pressure), we further examined the Component View by clicking on the L1 cache widget title. In the Component View, we identified long gaps between tasks, indicating the L1 cache is idle for most of the time. After this, we attempted to make sense of why the L1-Cache tasks did not start earlier. To reason about the process, we selected the second task to enable the Task View. We kept using the ``parent task'' bar in the Task View to trace back to the root of the tree.

\definecolor{wavefront_color}{RGB}{166,219,177}

As we traced back to the Compute Unit, we could check the memory instructions that triggered the L1 cache access. We could identify that the reason why the L1 did not have enough tasks was that the memory instructions were executed infrequently. Ideally, each Compute Unit can run 40 wavefronts \colorRect{wavefront_color} (as shown in \autoref{fig:case_study}(d)), so that there should be a large probability that at least one wavefront issues a memory instruction. However, in \autoref{fig:case_study}(d), we can see that only 3-5 wavefronts were executing in parallel. These wavefronts might be all executing compute instructions or waiting for existing memory accesses to complete, so they cannot generate new memory accesses. After making sense of the performance issue, we concluded that the root cause might lie in the Compute Unit that was not executing enough wavefronts. The global performance bottleneck was that the Command Processor was not running fast enough to dispatch work-groups. 

A solution to the problem above is to increase the dispatching speed. As we change the dispatching speed from one work-group per cycle to two work-groups per cycle, MGPUSim reported that the GPU execution time is reduced from $\approx 51ms$ to $\approx 44ms$, speeding up the execution by $1.16 \times$. If we compare the work-group and wavefront execution in \autoref{fig:case_study} (e) and (f), we can see that the Compute Unit had been executing more work-groups at the same time (see the 6th row). Overall, the findings from Daisen, and the minor modification of the GPU design, improved device utilization, and reduced the overall execution time.
 
\section{Evaluation}
 
To further evaluate Daisen, we conducted user studies with 10 GPU hardware designers (9 of the 10 also submitted the survey mentioned above; none of the 10 are authors of this paper). The average time they had worked in performance analysis was 4.6 years ($sd=2.8$). The highest degree of education among participants was split equally between master’s ($n=4$) and doctoral degrees ($n=4$), with one bachelor’s and one associate's degree. 
 
\subsection{Methodology}
 
We conducted our evaluation study remotely using online communication tools (e.g., Zoom). We asked participants to share their screens to allow us to observe how they interacted with the visualization framework. The evaluation study includes five parts. (1) Participants were provided with a walk-through tutorial. (2) They were asked to explore the tool on their own using the ``think-aloud'' method~\cite{lewis1982using} that encourages verbalizing their thoughts as they move through the visualization. In this process, participants were asked to explore a Matrix Transpose benchmark (1024$\times$1024 matrix, a well-known benchmark for GPU computing) running on an R9 Nano GPU. The dataset has a similar scale as compared to the dataset used in the case study. (3) They were asked to identify any performance bottlenecks they examined. (4) We conducted follow-up semi-structured interviews about their experience interacting with Daisen, accessing ease-of-use, what users liked and disliked about Daisen, and what changes they would make to help them identify performance issues more efficiently. (5) We asked each participant to fill out a post-study survey (see \autoref{fig:postsurvey} and supplemental materials for details) regarding how well Daisen facilitates performance analysis.  The whole session lasted about 90 minutes for each participant. 

\subsection{Findings}

Our participants gave overall positive feedback on Daisen, especially how Daisen can be used to help them identify and make sense of performance issues.
During the evaluation, 9 out of 10 participants successfully identified that the benchmark was a memory-intensive benchmark. 7 out of 10 participants pointed out that reducing the latency or increasing the bandwidth of the caches and the memory controllers can improve performance.

\boldsec{Value of visualizing GPU traces in Daisen. }
Participants appreciated the GPU simulation visualizations and some expressed excitement about using Daisen to examine GPU performance issues.
They also described how they typically visualize execution information. 
Half indicated that they first added ``print functions'' in the GPU simulator to get more performance information, exported to a CSV file and then visualizing the trace data separately (in Python) to explore and identify performance issues. 
E.g., P03 said \textit{``My debugging is using prints to get any of these numbers. But is also cumbersome to either have another script to process those files and then read them.''} 
P03 further appreciated Daisen's capability of providing visual clues related to performance issues.
P03 said \textit{``Daisen gives a pretty good picture of this benchmark is suffering because of certain factors. Then I can take a design step to improve something.''} 
P07 also liked how Daisen supports visualizing GPU traces, comparing it with VTune~\cite{reinders2005vtune} and nvprof~\cite{nvprof}, other trace visualization tools. 
P07 mentioned that \textit{``nvprof also gives you some kind of views but can only show which kernels are running (from a high-level and coarse-grained perspective). However, Daisen enables users to understand which portion and phases of the application are getting low [performance] and have problems.'' }


\boldsec{Overview allows quick hotspot identification.} 7 out of 10 participants started to use the filter box to narrow down the components after browsing no more than 4 pages. None of the participants attempted to check all pages. Moreover, 9 out of 10 participants displayed little difficulty finding the desired components by typing in commonly-used computer architecture terms such as ``L1'' or ``TLB''. The participants' way of interacting with the Overview Panel suggests our participants (as domain experts) can easily learn how to use Daisen to solve problems.

Participants also frequently zoomed in and out to search hot spots that took longer in execution.  P05 specifically valued being able to browse behaviors of various components, as \textit{``it gives the users the ability to verify their hypothesis on different applications and see which resources cause the constraint in the system.''}

\boldsec{Task View enables navigation between components. }
Our participants used the Task View mainly to navigate across components and explore execution relationships between the Task View and the Component View. P04 and P08 particularly liked the Task View's ability to navigate between components. P08 said \textit{``It is really really nice to have this type of navigation between tasks and subtasks.''}

\boldsec{Component View helps examine low-level behaviors. }
Our participants mainly used the Component View to get a more in-depth understanding of the performance of a particular component, especially when they saw conflicting things happening in an overview. For example, P03 recalled a situation where he needed a detailed view. He explained that when he tried to understand what instructions the workload was executing, he needed more evidence to support his hypothesis on how to improve the design with the help of the Component View. Similarly, P01 specifically mentioned how well Daisen helped him understand simulation behavior as compared with some profilers he had used before. He commented that existing profiling tools failed to provide sufficient details for making sense of performance issues.

Though all participants foresee themselves using Daisen to facilitate their research in GPU hardware design and computer architecture research, they suggested a few things that can be improved to further help them examine GPU performance issues. These features include (1) educational support and tutorings, such as automatic screen recording and more user guidance, and (2) integration of summary statistics into Daisen to enable users with various levels of visualization literacy to understand the performance (e.g., statistically summarizing 70\% of the instructions being executed).

\begin{figure}[tb]
 \centering
 \includegraphics[width=\linewidth]{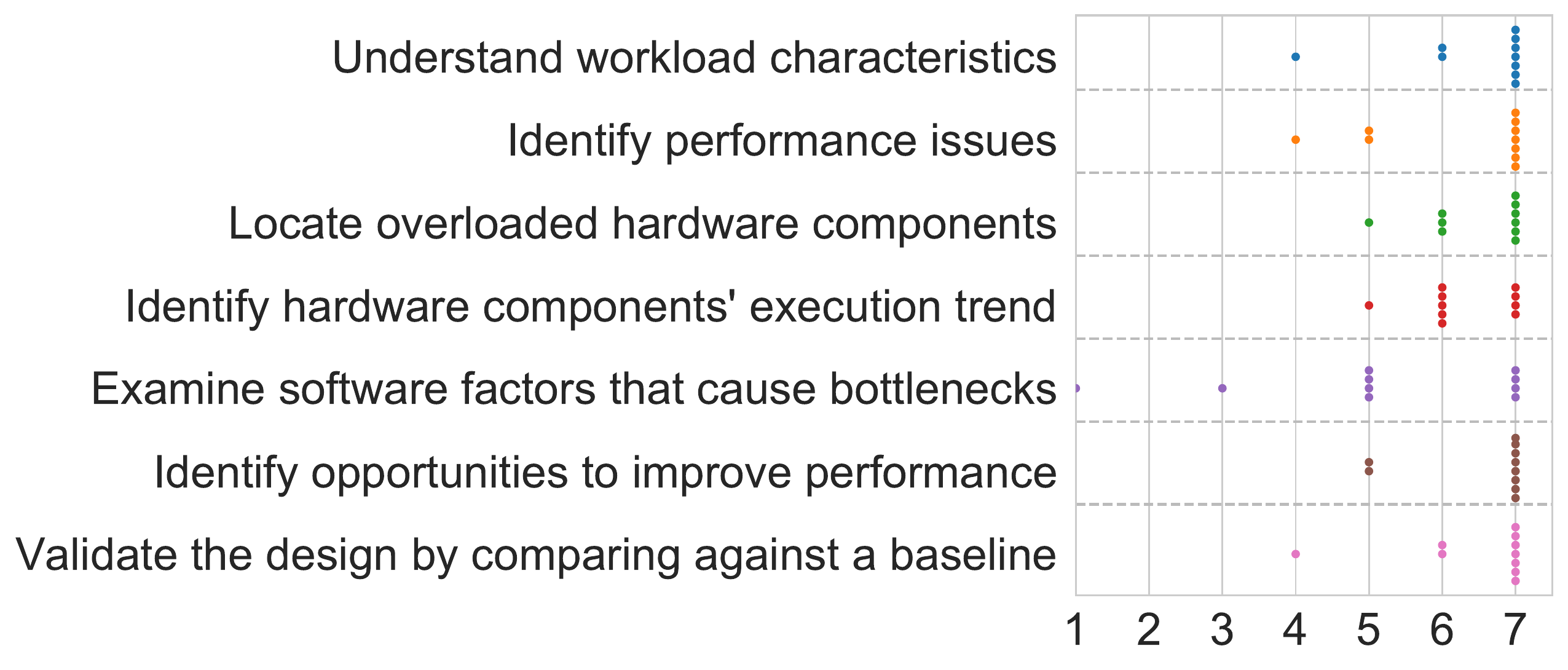}
 \caption{Post-study survey results: Participants rated whether they were able to perform a set of tasks using a 7-point scale, labeled from Strongly Disagree (1) to Strongly Agree (7).} 
 \label{fig:postsurvey}
 \vspace{-0.8cm}
\end{figure}

\boldsec{Post-study survey results. }
Overall, as shown in \autoref{fig:postsurvey}, participants found that using Daisen helped them 
understand key characteristics of workloads (median 7, IQR 0.75), identify performance issues (median  7, IQR 1.5), 
locate overloaded hardware components (median 7, IQR 1),  explore trends in executions (median 6, IQR 1), 
examine software-related factors that cause the performance bottlenecks (median 5, IQR 2), identify opportunities to improve performance (median 7, IQR 0), and
validate the new design by comparing with a baseline (median 7, IQR 0.75). 

\section{Discussion}

Participant feedback about Daisen was very positive overall. In this section, we reflect on our task analysis, design, and evaluation.

Our task analysis and abstraction was drawn from our survey and interview studies with GPU designers. As we described, some tasks are particularly essential for GPU performance analysis, such as comparing metrics of parallel components (Task 2.3.2). Some other tasks may be applicable to other sibling domains (e.g., designing CPUs and domain-specific accelerators); thus, we suggest future research to adapt the Daisen design to fit design requirements for other types of digital circuit devices.

\boldsec{Generalization vs. specification.} One challenge we encountered when designing Daisen was to balance between generalization and specialization. On the one hand, we can pursue generalizability that aims to benefit a wider range of audiences~\cite{thudt2017expanding}. On the other, design studies have focused on seeking solutions for a particular problem domain~\cite{munzner2008process}.  
Daisen's design avoids a close binding between the visualization style and the component to be examined. Instead, we use the same views to reveal the behavior of most of the components on a GPU, such as Compute Units, caches, and memory controllers.



Our effort to maintain a general framework created some challenges.
Our participants felt that the component sorting in the Overview Panel was not intuitive, and that it took some time for them to understand how  components were connected.
Although most participants indicated that they were able to infer the component-specific metrics from the component-independent metrics, some (P03, P07) also wanted to see traditional component-specific metrics to help them understand the execution. Therefore, more work is needed to examine the best trade-off between generalization and specialization. For example, we plan to allow users to tag the tasks with user-specified properties and use the tags to visualize component-specific metrics.

\boldsec{GPU visualization for non-expert users.} We noticed that more experienced GPU designers more quickly grasped key concepts visualized in Daisen. This shortcoming raises a question for future research: how to design a visualization environment for users with different levels of expertise and help overcome the steep learning curve. 
For example, non-experts may not be able to pinpoint interesting components in the Overview Panel and may end up browsing pages. 
Many approaches have been explored to help users comprehend how visualization works, such as step-by-step wizards, forums, and animated video tutorials~\cite{fernquist2011sketch, grossman2010toolclips, matejka2011ambient, pongnumkul2011pause}. 
In addition to these strategies, we plan to add better navigation support (e.g., grouping, ordering) in the Overview Panel and provide visual hints about how the components are connected. We also plan to have each component specify the expected metric values, making it easier for beginners to compare metrics.


\boldsec{Limitations.} There are several limitations in our current design which we are planning to improve.
While abstracting hardware execution behavior can represent most digital circuit behaviors, some corner cases may not perfectly apply.
E.g., DRAMs usually spontaneously and periodically initiate a refresh action, which does not belong to the task hierarchy tree.
We currently do not capture such behavior but will generalize to a graph data representation.
Another limitation is that we do not highlight communication between GPU hardware components.
In chip design, on-chip network bandwidth is a scarce resource and can limit overall performance.
We plan to add a network view and time-lapse traffic visualization.
Moreover, we currently do not support comparing executions between different benchmark executions and different hardware configurations.
Finally, the current design requires a good knowledge of computer architecture and information about the simulation being visualized.
Future work also includes using heuristic algorithms as well as narrative solutions to highlight potential bottlenecks.

\section{Conclusion}

Visualizing massively parallel GPU hardware execution has been challenging due to the large number of digital circuit components and the complex interplay between them. To address this problem, we have described Daisen, which includes a dedicated data representation for simulator-generated hardware executions and a web-based interactive visualization tool. Using a case study and user evaluation studies, we demonstrated how Daisen can be used to help GPU hardware designers identify performance bottlenecks. Daisen is currently under active development, with more features and improvements being added to the tool. We hope the design of Daisen can serve as the first step of a series of works that can help hardware designers build explainable computer architecture, since only human-understandable computer architecture can avoid costly hardware bugs, performance bottlenecks, reliability issues, and security vulnerabilities.



\section*{Acknowledgments}
We thank our reviewers for their constructive feedback. We also thank AMD for supporting this work. 

\balance
\bibliographystyle{eg-alpha-doi}  
\bibliography{ref}        


\newpage

\end{document}